# On the structure of SbTeI

R. Sereika[1,*], R. Žaltauskas[1], Š. Varnagiris[2], M. Urbonavičius[2], F. Liu[3], Y. Ding[3], D. Milčius[1,2]

[1]*Vytautas Magnus University, K. Donelaičio Str. 58, LT-44248 Kaunas, Lithuania*

[2]*Center for Hydrogen Energy Technologies, Lithuanian Energy Institute, LT-44403 Kaunas, Lithuania*

[3]*Center for High Pressure Science and Technology Advanced Research, Beijing 100094, China*

**Abstract**

**Antimony telluroiodide (SbTeI) is predicted to be a promising material in many technological applications based on theoretical simulations, however the bulk structure solution remains elusive. We consolidate SbTeI belonging to the base-centered monoclinic lattice with a space group C 2/m by combining single crystal X-ray diffraction and X-ray photoemission spectroscopy techniques. The atomic arrangement of the reported crystal structure is remarkable with one-dimensional double-chains forming two-dimensional blocks. In this structure, the $Sb^{3+}$ ion is surrounded by $Te^{2-}$ and $I^-$, which is distinguishable by an incomplete polyhedron resulting in the $5s^2$ (Sb) lone pair electrons in the valence band. Manipulation of this material with pressure to induce novel structures and properties is highly anticipated.**

*\*Corresponding author: <raimundas.sereika@vdu.lt>*

# I. Introduction

Antimony and bismuth chalcohalides show great promise for photovoltaic applications [1-5], where they also feature flexible electronics and novel piezoelectric devices [6-8]. In contrast to common Pb and Sn, compounds containing trivalent Sb/Bi have ecological advantage and demonstrate good stability parameters in air [9]. Recently, SbTeI material was highlighted as relevant absorbent of the solar spectrum with high conversion efficiency evaluated from the theoretical absorption coefficients [10]. In addition, it shows low thermal conductance of the sheet which is favorable for 2D thermoelectric materials from the monolayer [11]. Furthermore, Rashba effect in a single-layer antimony telluroiodide was predicted for potential spintronics applications [12]. Theoretical calculations of the low formation energy and real phonon modes imply that the system is stable. It should be noted that the Rashba spin splitting in the spin-orbit coupling (SOC) band structure of single-layer SbTeI is significantly larger than that of a number of two-dimensional systems, including surfaces and interfaces.

A lot of research has been done recently with the relatively close compound BiTeI, which has many unique properties, including the Rashba effect. [13, 14]. However, structurally, SbTeI is much different from BiTeI which is clearly layered 2D material and of significantly higher symmetry – trigonal system, space group $P3m1$ (No. 156) [15]. As a semiconductor, SbTeI has more in common with 1D ferroelectric SbSI-type compounds, most of which are characterized by orthorhombic crystal symmetry comprising of infinite double-chains [1-7]. Yet, in this work, we show that SbTeI may be an intermediate example between 1D – 2D systems, as it holds somewhat both layer and chain arrangement.

In the literature, there are only a few studies dealing with crystal structure of SbTeI and their results are quite different. This causes confusion, for example, when performing theoretical *ab-initio* calculations [10, 16-18]. Initially, it was announced that the structure may adopt the orthorhombic or even the lowest triclinic symmetry [19-21]. Investigating more precisely, A. G. Papazoglou and P. J. Rentzeperis found that the symmetry should be monoclinic with the space group *C* 2/*m* [22]. On the other hand, theoretical simulations indicate that bulk SbTeI with the trigonal symmetry (analogous to the BiTeI crystal) exhibits energy 19 meV/atom lower than that of monoclinic SbTeI [12]. Therefore, seeing the growing interest in these ternary materials as well as many uncertainties, we have undertaken growth and structural analysis of pure single crystals of SbTeI. In this study, we also examine their surface chemical composition and internal electronic properties.

## II. Methods

SbTeI compound was prepared by heating a high purity mixture of antimony (99.999%), tellurium (99.999%), and iodine (≥99.99%) purchased from "Sigma-Aldrich". Stoichiometric amounts of Sb, Te, and I powders were sealed in an evacuated quartz tube and placed in a rotating furnace. The temperature was slowly raised up to 670 K and held at this level for 24 h. Then, the tube was fixed and kept stable in turn temperature, which was gradually lowered to the room temperature. The resulting crystals were black-gray in color and had a mirror-like surface (see Fig. 1a).

A specimen of SbTeI was used for the X-ray crystallographic analysis. The X-ray intensity data were measured on a Bruker D8 VENTURE PHOTON II system equipped with a Microfocus Incoatec Ims 3.0 (Ag Kα, $\lambda$ = 0.56086 Å) and a multilayer optic monochromator. A total of 872 frames were collected. The total exposure time was 0.48 h. The frames were integrated with the Bruker SAINT software package using a narrow-frame algorithm. The integration of the data using a monoclinic unit cell yielded a total of 2132 reflections to a maximum $\theta$ angle of 22.12° (0.74 Å resolution), of which 604 were independent (average redundancy 3.530, completeness = 100.0 %, $R_{int}$ = 4.36 %, $R_{sig}$ = 3.97 %) and 542 (89.74%) were greater than $2\sigma(F^2)$. The final parameters listed in the Tables 1-4 are based upon the refinement of the XYZ-centroids of 2513 reflections above 20 $\sigma(I)$ with 4.476° < 2θ < 44.15°. The data were corrected for absorption effects using the Multi-Scan method (SADABS). The ratio of minimum to maximum apparent transmission was 0.462. The structure was solved and refined using the Bruker SHELXTL Software Package, using the space group $C2/m$ (No. 12, unique axis $b$), with $Z$ = 4 for the formula unit, SbTeI. The final anisotropic full-matrix least-squares refinement on $F^2$ with 19 variables converged at $R_1$ = 3.27% for the observed data and $wR_2$ = 8.72 % for all data. The goodness-of-fit was 1.20. The largest peak in the final difference electron density synthesis was 1.792 e$^-$/Å$^3$ and the largest hole was -1.711 e$^-$/Å$^3$ with an RMS deviation of 0.387 e$^-$/Å$^3$. On the basis of the final model, the calculated density was 6.009 g/cm$^3$ and F(000), 624 e$^-$.

X-ray photoelectron spectroscopy (XPS) was used to investigate the surface composition and chemical state of the elements within a SbTeI. X-ray photoelectron spectra were acquired at room temperature by using a PHI Versaprobe 5000 spectrometer. The photoelectrons were excited using monochromatized 1486.6 eV Al radiation, 25 W beam power, 100 μm beam size and 45° measurement angle. Sample charging was compensated using dual neutralization system consisting of low energy electron beam and ion beam. The random C 1$s$ line with binding energy fixed at 284.6 eV was used for correction of the charging effects. After background subtraction, a non-linear least squares curve fitting routine with a Gaussian/Lorentzian product function was used for the analysis of XPS spectra.

Theoretical simulations of different structures and compositions within the ternary Sb-Te-I system were performed using the crystal structure prediction method implemented in the USPEX code [23, 24]. The USPEX was used in coupling with the first-principles calculations within the framework of density functional theory (DFT) using the Vienna *ab initio* simulation package VASP [25]. For the determinations of the total energies, optimized lattice structures and their corresponding electronic structures, the Perdew, Burke and Ernzerhof generalized gradient exchange-correlation functional (PBE-GGA) was selected [26]. The plane–wave kinetic energy cutoff was set to 500 eV and a Brillouin zone sampling resolution was $2\pi \times 0.05$ Å$^{-1}$. All structures were relaxed at ambient pressure and 0 K, and the enthalpy was used as fitness function. In addition, for the structural candidates searched by USPEX, we further performed DFT analysis using the WIEN2k software [27]. Here, the calculations were carried out on a mixed basis set of augmented plane waves (APW) and linearized augmented plane waves (LAPW). The exchange and correlation potentials have been treated according to the PBE-GGA scheme. Wave functions in the interstitial regions were expanded in plane waves, with the plane wave cutoff chosen so that $R_{MT}k_{max} = 7$ (where $R_{MT}$ represents the smallest atomic sphere radius and $k_{max}$ is the magnitude of the largest wave vector). The energy separation between core and valance states was -6.0 Ry. For the structure optimization, we required that the forces are smaller than 1 mRy/a.u. When performing self-consistent calculations for fixed geometry, the iteration halted when the difference in the eigenvalues was less than 0.0001 between steps of convergence criterion.

## III. Results and discussion

Our single crystal X-ray diffraction (XRD) measurements indicate that SbTeI adopts a monoclinic lattice (space group *C* 2/*m*) with the unit cell parameters $a$ = 13.698(6) Å, $b$ = 4.2289(18) Å, $c$ = 9.191(4) Å, $β$ =128.626° (see more details provided in Table 2 and supplementary material). This coincides with the results presented in Ref. 21, where the unit cell parameters $a$ = 13.7008(10) Å, $b$ = 4.2418(3) Å, $c$ = 9.2005(8) Å, and $β$ =128.631° were reported. This also confirms that SbTeI crystals synthesized at ambient pressure are indeed monoclinic rather than orthorhombic or trigonal, in contrast to the triclinic symmetry ($P1$, $P\bar{1}$) announced in Ref. [20].

As it was briefly mentioned in the introduction, the double-chains of SbTeI extending through the shorted *b*-axis of the unit cell are very similar to those in SbSI-type structure. However, double-chains of SbTeI are uniquely situated forming plate-like blocks (see Fig. 1b). These blocks are likely held together by a weak van der Waals connection. Such an arrangement has strong anisotropic properties and a rather brittle nature, which can be easily observed by mechanical pressing of the sample. The calculated universal anisotropy index, $A^U$, gives the value of 1.85 indicating strong anisotropy as well [18]. For comparison, BiTeI has $A^U$ = 0.37 and $A^U$ = 0 represents locally isotropic crystals. As for the atomic bonds inside the chains, considerable covalent character can be pondered in the Sb-Te bond because it is shortest in length and closer to the sum of the covalent radii [22]. Overall, a mixed covalent-ionic-metallic character prevails the structure where metallicity is gained from tellurium. Mössbauer spectroscopy showed that Sb systems containing tellurium, in most cases, have negative quadrupolar splitting, indicating a tendency to metallic character [28-31]. On the contrary, sulfur (as well as selenium in similar systems) usually tends to ionic character.

The metallicity of bonds as well as forbidden gap value are closely related to the stereochemical activity of the lone pair which is often considered as a key factor for the efficient carrier transport in optoelectronic materials. On the other hand, the example of SbSI showed that the lone pair of antimony plays an active role in the stabilization of ferroelectricity [32]. The monoclinic SbTeI variation indeed has the lone pair of antimony. Figure 1c shows that there is one missing atom at one edge of the octahedron for SbTeI. Thus, the octahedron is incomplete and takes square pyramidal shape: tellurium is at the apex of a pyramid whose base is one of the $Te_2I_2$ faces of the triangular prism. Although lone pair does not participate in bonding, its properties are connected to the local atomic arrangement of antimony. In fact, the stereochemical activity of the lone pair is sensitive to the local environment because it affects the loss of sphericity of the 5*s'* electron distribution around the Sb atom. Usually, small coordination polyhedrons with strong covalent bonds have active lone pairs and their stereochemical activity increases with the distortion of the local environment. The structural modifications then correspond to changes in the nature of bonds from more or less ionic to covalent or metallic. On the basis of chemical bonding, the stereochemical activity of the lone pair for SbTeI is considered to be low in activity [31]. However, it is likely that chemical or physical pressure-induced changes would cause electronic transformations because the lone pair electrons are directed to occupy the space between plate-like blocks. Such an effect on the structure can be quite significant, as in the case of layered $Bi_2O_2S$, where the high-pressure enforces the lone pair electrons to disappear and triggers 2D-to-3D structural transition [33].

In general, to balance the charge, SbTeI with 1:1:1 stoichiometry should be formed from a trivalent cation with divalent and monovalent anions, respectively. Here, we used XPS methods to define the chemical state and related peculiarities of SbTeI. Figure 2 shows the peaks of Sb 3*d*,

I 3*d*, Te 3*d* and valence band (VB) spectrum collected at room temperature. The Sb 3*d* spin–orbid doublet is situated at the binding energies (BEs) of 530.3 eV and 539.7 eV, for Sb $3d_{5/2}$ and Sb $3d_{3/2}$, respectively. The fitting of the experimental data indicates that each line of the Sb 3*d* spin–orbit doublet has an additional minor component approximately spaced by 1.4 eV (see Fig. 2a). All four lines are well matching with Sb 3*d* in sonochemically prepared SbSI [34]. For the antimony chalcohalides the lines divided into two groups indicate separate states of surface and bulk [35-37]. Although the surface tends to oxidize, both BEs represent $Sb^{3+}$ species. No other antimony species contributions corresponding to $Sb^{4+}$ or $Sb^{5+}$, which appear at slightly higher BEs [38], are seen. The oxidation of the surface is also evident from I 3*d* and Te 3*d* spectra given in Figs. 2b and 2c, respectively. We observe that I 3*d* region has well separated spin-orbit doublet, $\Delta = 11.5$ eV. The position of I $3d_{5/2}$ is at BE = 618.8 eV and of I $3d_{3/2}$ – at BE = 630.3 eV. The chemical shift is -0.7 eV to lower energy values (pure I $3d_{5/2}$ it is at BE = 619.5 eV and for I $3d_{3/2}$ at BE = 631.0 [39]). The main lines have two adjacent components spaced 1.3 eV apart. The main Te $3d_{5/2}$ and $3d_{3/2}$ peaks were located at BEs of 573.1 and 583.5 eV, respectively. This indicates metallic tellurium similar, as for example, in $PtTe_2$ crystals [40]. In addition, rather strong peaks at BEs of 576.32 and 586.72 eV were observed as well. The later peaks can be assigned to the Te-O bonds which, as mentioned before, are often found on the surfaces of chalcogenide materials due to the rapid surface oxidation after exposure to the ambient atmosphere [40-42].

Figure 2b shows the XPS spectrum of the VB for the SbTeI crystals at room temperature. In the given region, three bands can be singled out. The shape of this structure is comparable to the VBs reported for the SbSI-type crystals [34, 35, 43] and corresponds well to the DFT simulated total density of states (T-DOS) (see Fig. 3a). It should be noted that such a comparison is only of a general nature, as there is a large temperature difference between the experimental measurements and the theoretical calculations [43]. However, the well separated bands allow their composition to be identified. According to the simulations, the lowest BE band is mainly the anion-*p*-dominated consisting of Sb 5*p*, I 5*p* and Te 5*p* hybridized orbitals (Fig. 3b). The partial densities of states (P-DOS) reveal only very small portion of the Sb 5*s* involved in this band (Fig. 3c). The second band should be related to the lone pair of antimony, because it has most pronounced 5*s* states here. Overall, the second and third bands originate from Sb 5*s*, I 5*s* and Te 5*s* hybridized orbitals. In addition, as P-DOS indicates, the conduction band (CB) composed of Sb 5*p*, I 5*p* and Te 5*p* hybridized orbitals where cation has the largest contribution.

Sb–Te–I system is known to have only one ternary compound SbTeI [44]. Our structure search for 1:1:1 composition using the evolutionary algorithm implemented in USPEX software predicts a series of candidates to match the experiment. The lowest enthalpy structures were taken into account and their corresponding energy–volume curves are presented in Fig. 4 (with more details provided in the supplementary material). Here, the volume is given in terms of the reduced volume, $V/V_0$, emphasizing that considered systems have no points of contact at a given range. The resulting $E = f(V)$ curves shown in Fig. 4 correspond to fitting of the calculated points using a Birch-Murnaghan expression [45, 46]:

$$E(V) = E_0 + \frac{9}{8} V_0 B_0 \left[ \left(\frac{V_0}{V}\right)^{\frac{2}{3}} - 1 \right]^2 + \frac{9}{16} B_0 (B' - 4) V_0 \left[ \left(\frac{V_0}{V}\right)^{\frac{2}{3}} - 1 \right]^3, \qquad (1)$$

where $E_0$, $V_0$, $B_0$ and $B_0'$ are the equilibrium energy, the volume, the zero pressure bulk modulus and its pressure derivative, respectively. The derived value of the bulk modulus characterizes

SbTeI as soft material, $B_0 \approx 38$ GPa. This value is also in a good agreement with the data for ternary bismuth tellurohalides [18, 47].

A series of energy calculations with fixed symmetry determine that orthorhombic *Pnam* is energetically unfavorable but two other phases (trigonal *P*3*m*1 and monoclinic *Cm*) justified as suitable and are more stable than experimentally observed *C* 2/*m*. In all cases, SbTeI is found as semiconductor where the forbidden gap is surrounded by intertwined *p* orbitals but, interestingly, for the lowest energy phases, it has clearly layered 2D structure with octahedron being complete (see supplementary material). The absence of both the inversion symmetry and out-of-plane mirror symmetry makes this compound much attractive. However, to obtain bulk SbTeI experimentally in *P*3*m*1 or *Cm* form is likely to be difficult process requiring very special high-pressure-temperature conditions or even impossible. Despite the technical challenges, other forms such as Janus monolayer are intriguing as they can induce novel electronic and piezoelectric properties as well. Furthermore, the application of external pressure, which is an alternative thermodynamic parameter, could modulate the atomic and electronic structures by changing the bond distances. In this regard, SbTeI may have a transition from the monoclinic – lone-pair phase to the phase of completed polyhedron.

## IV. Conclusions

Single crystal XRD measurements indicate that SbTeI obtained by using standard synthesis methods at ambient pressure are of base-centered monoclinic lattice with a space group *C* 2/*m*. The structure consists of double-chains that lie in plate-like blocks. Therefore, these crystals are characterized by strong anisotropy, while XPS shows their surface in the air is being oxidized. Such atomic layout is considerably different from BiTeI compound that settle in trigonal *P*3*m*1 at ambient. Despite the fact that bismuth and antimony are both trivalent cations, their ternary compounds have several differences that may be detected in this fashion, such as SbTeI being made up of unfinished octahedra, having a structure that is closer to the 1D arrangement, and having a larger band gap. The valence band spectra observed for SbTeI indicated three bands that are dominated by *p* and *s* orbital hybridization while the conduction band is composed only of hybridized *p* orbitals. Such hybridization should result in strong optical absorption near band gap energies, which is important for a solar absorber material. Another important factor is the lone pair which plays a key role in the SbTeI structure and could be further manipulated with pressure to generate novel structures and electrical, and optical properties.

**Supplementary Material**

Detailed crystallographic information on the structure of SbTeI (Crystallographic Information File "cif") determined experimentally and predicted theoretically can be found in the supplementary material.

**Data Availability Statement**

The data that supports the findings of this study are available within the article and its supplementary material.

**REFERENCES**

[1] H. Kunioku, M. Higashi, R. Abe, Low-Temperature Synthesis of Bismuth Chalcohalides: Candidate Photovoltaic Materials with Easily, Continuously Controllable Band gap, Sci. Rep. 6 (2016) 32664.
[2] R. Nie, J. Im, S. I. Seok, Efficient Solar Cells Employing Light-Harvesting $Sb_{0.67}Bi_{0.33}SI$, Adv. Mater. 31 (2019) 1808344.

[3]  A. M. Ganose, S. Matsumoto, J. Buckeridge, D. O. Scanlon, Defect Engineering of Earth-Abundant Solar Absorbers BiSI and BiSeI, Chem. Mater. 30 (2018) 3827−3835.

[4]  K. Mistewicz, M. Nowak, D. Stróż, A Ferroelectric-Photovoltaic Effect in SbSI Nanowires, Nanomaterials 9 (2019) 580.

[5]  S. Inagaki, M. Nakamura, H. Hatada, R. Nishino, F. Kagawa, Y. Tokura, M. Kawasaki, Growth of visible-light-responsive ferroelectric SbSI thin films by molecular beam epitaxy, Appl. Phys. Lett. 116 (2020) 072902.

[6]  Y. Purusothaman, N. R. Alluri, A. Chandrasekhar, S.-J. Kim, Photoactive piezoelectric energy harvester driven by antimony sulfoiodide (SbSI): A $A_VB_{VI}C_{VII}$ class ferroelectric-semiconductor compound, Nano Energy 50 (2018) 256–265.

[7]  D. Tiwari, F. Cardoso-Delgado, D. Alibhai, M. Mombrú, D. J. Fermín, Photovoltaic Performance of Phase-Pure Orthorhombic BiSI Thin-Films, ACS Appl. Energy Mater. 2 (2019) 3878−3885.

[8]  S.-D. Guo, X.-S. Guo, Z.-Y. Liu, Y.-N. Quan, Large piezoelectric coefficients combined with high electron mobilities in Janus monolayer XTeI (X = Sb and Bi): A first-principles study, J. Appl. Phys. 127 (2020) 064302.

[9]  R. Nishikubo, H. Kanda, I. García-Benito, A. Molina-Ontoria, G. Pozzi, A. M. Asiri, M. K. Nazeeruddin, A. Saeki, Optoelectronic and Energy Level Exploration of Bismuth and Antimony-Based Materials for Lead-Free Solar Cells, Chem. Mater. 32 (2020), 6416−6424.

[10] C. Tablero, Optical properties of Sb(Se,Te)I and photovoltaic applications, J. Alloy. Compd. 678 (2016) 18−22.

[11] S.-D. Guo, A.-X, Zhang, H.-C. Li, Potential 2D thermoelectric material *A*TeI (*A* = Sb and Bi) monolayers from a firstprinciples study, Nanotechnology 28 (2017) 445702.

[12] H. L. Zhuang, V. R. Cooper, H. Xu, P. Ganesh, R. G. Hennig, P. R. C. Kent, Rashba effect in single-layer antimony telluroiodide SbTeI, Phys. Rev. B 92 (2015) 115302.

[13] Y. Qi, W. Shi, P. G. Naumov, N. Kumar, R. Sankar, W. Schnelle, C. Shekhar, F.-C. Chou, C. Felser, B. Yan, S. A. Medvedev, Topological Quantum Phase Transition and Superconductivity Induced by Pressure in the Bismuth Tellurohalide BiTeI, Adv. Mater. 29 (2017) 1605965.

[14] X. Li, Y. Sheng, L. Wu, S. Hu, J. Yang, D. J. Singh, J. Yang, W. Zhang, Defect-mediated Rashba engineering for optimizing electrical transport in thermoelectric BiTeI, npj Comp. Mater. 6 (2020) 107.

[15] A. Shevelkov, E. Dikarev, R. Shpanchenko, B. Popovkin, Crystal Structures of Bismuth Tellurohalides BiTe*X* (*X* = Cl, Br, I) from X-Ray Powder Diffraction Data, J. Solid State Chem. 114 (1995) 379−384.

[16] T. Ozer, S. Cabuk, Ab initio study of the lattice dynamical and thermodynamic properties of SbXI (X=S, Se, Te) compounds, Comput. Condens. Matter 16 (2018) e00320.

[17] T. Ozer, S. Cabuk, First-principles study of the structural, elastic and electronic properties of SbXI (X=S, Se, Te) crystals, J. Mol. Model 24 (2018) 66.

[18] H. Koc, S. Palaz, A. M. Mamedov, E. Ozbay, Optical, electronic, and elastic properties of some $A^5B^6C^7$ ferroelectrics (A=Sb, Bi; B=S, Se; C=I, Br, Cl): First principle calculation, Ferroelectrics 511 (2017) 22–34.

[19] Von E. Dönges, Über Chalkogenohalogenide des dreiwertigen Antimons und Wismuts. III. Über Tellurohalogenide des dreiwertigen Antimons und Wismuts und über Antimon-und Wismut(III)-tellurid und Wismut(III)-selenid, Z. Anorg. Allg. Chem. 265 (1951) 56−61.

[20] A. Ibanez, J.-C. Jumas, J. Olivier-Fourcade, E. Philippot, M. Maurin, Sur les chalcogeno-iodures d'antimoine SbXI(X =S, Se, Te): Structures et spectroscopie Mössbauer de $^{121}$Sb, J. Solid State Chem. 48 (1983) 272−283.

[21] O. Madelung. Semiconductors: data handbook. Springer-Verlag, Berlin Heidelberg, pp 664–673 (2004).

[22] A. G. Papazoglou and P. J. Rentzeperis, The crystal structure of antimony telluroiodide, SbTeI, Z. Kristallogr. 165 (1983) 159−167.

[23] A. O. Lyakhov, A. R. Oganov, H. T. Stokes, Q. Zhu, New developments in evolutionary structure prediction algorithm USPEX, Comp. Phys. Comm. 184 (2013) 1172–1182.

[24] A. R. Oganov, A. O. Lyakhov, M. Valle, How Evolutionary Crystal Structure Prediction Works–and Why, Acc. Chem. Res. 44 (2011) 227–237.

[25] G. Kresse, J. Furthmller, Efficiency of ab-initio total energy calculations for metals and semiconductors using a plane-wave basis set, Comput. Mater. Sci. 6 (1996) 15–50.

[26] J. P. Perdew, K. Burke, M. Ernzerhof, Generalized Gradient Approximation Made Simple, Phys. Rev. Lett. 77 (1996) 3865.

[27] P. Blaha, K. Schwarz, G. K. H. Madsen, D. Kvasnicka, J. Luitz, Wien2k, An Augmented Plane Wave plus Local orbital Program for Calculating the Crystal Properties, 2001.

[28] A. Ibanez, J. Olivier-Fourcade, J. C. Jumas, E. Philippot, M. Maurin, Relations Structures-Propriétés Physiques dans quelques Semiconducteurs à Paire Électronique non liée, Z . anorg. allg. Chem. 540 (1986) 106−116.

[29] I. Lefebvre, M. Lannoo, G. Allan, L. Martinage, Theoretical Mössbauer isomer shift of antimony chalcogenides, Phys. Rev. B 38 (1988) 8593.

[30] I. Lefebvre, G. Allan, M. Lannoo, J. Olivier-Fourcade, J. C. Jumas, M. Maurin, Electronic structure of unconventional antimony chalcogenides: Theoretical calculations and $^{121}$Sb Mössbauer spectroscopy, Hyperfine Interact. 53 (1990) 351−354.

[31] J. Olivier-Fourcade, A. Ibanez, J. C. Jumas, M. Maurin, I. Lefebvre, P. Lippens, M. Lannoo, G. Allan, Chemical Bonding and Electronic Properties in Antimony Chalcogenides, J. Solid State Chem. 87 (1990) 366−377.

[32] D. Amoroso, S. Picozzi, Ab initio approach to structural, electronic, and ferroelectric properties of antimony sulphoiodide, Phys. Rev. B 93 (2016) 214106.

[33] K. Bu, H. Luo, S. Guo, M. Li, D. Wang, H. Dong, Y. Ding, W. Yang, X. Lü, Pressure-Regulated Dynamic Stereochemical Role of Lone-Pair Electrons in Layered $Bi_2O_2S$, J. Phys. Chem. Lett. 11 (2020) 9702−9707.

[34] M. Nowak, E. Talik, P. Szperlich, D. Stróż, XPS analysis of sonochemically prepared SbSI ethanogel, Appl. Surf. Sci. 255 (2009) 7689–7694.

[35] J. Grigas, E. Talik, M. Adamiec, V. Lazauskas, X-ray photoelectron spectra and electronic structure of quasi-one-dimensional SbSeI crystals, Condens. Matter. Phys. 10 (2007) 101–110.

[36] L. Santinacci, G. I. Sproule, S. Moisa, D. Landheer, X. Wu, A. Banu, T. Djenizian, P. Schmuki, M. J. Graham, Growth and characterization of thin anodic oxide films on n-InSb(1 0 0) formed in aqueous solutions, Corros. Sci. 46 (2004) 2067–2079.

[37] H. Bryngelsson, J. Eskhult, L. Nyholm, M. Herranen, O. Alm, K. Edström, Electrodeposited Sb and $Sb/Sb_2O_3$ Nanoparticle Coatings as Anode Materials for Li-Ion Batteries, Chem. Mater. 19 (2007) 1170–1180.

[38] R. Izquierdo, E. Sacher, A. Yelon, X-ray Photoelectron spectra of antimony oxides, Appl. Surf. Sci. 40 (1989) 175–177.

[39] C. D. Wagner, J. F. Moulder, L. E. Davis, W. M. Riggs, Handbook of X-ray Photoelectron Spectroscopy (Perkin–Elmer Corporation, Physical Electronics Division, 1995).

[40] A. Politano, G. Chiarello, C.-N. Kuo, C. S. Lue, R. Edla, P. Torelli, V. Pellegrini, D. W. Boukhvalov, Tailoring the Surface Chemical Reactivity of Transition-Metal Dichalcogenide $PtTe_2$ Crystals, Adv. Funct. Mater. 28 (2018) 1706504.

[41] H. Bando, K. Koizumi, Y. Oikawa, K. Daikohara, V. A. Kulbachinskii, H. Ozaki, The time-dependent process of oxidation of the surface of $Bi_2Te_3$ studied by x-ray photoelectron spectroscopy, J. Phys.: Condens. Matter 12 (2000) 5607–5616.

[42] A. A. Volykhov, J. Sánchez-Barriga, A. P. Sirotina, V. S. Neudachina, A. Frolov, E. A. Gerber, E. Yu. Kataev, B. Senkovsky, N. O. Khmelevsky, A. A. Aksenenko, N. V. Korobova, A. Knop-Gericke, O. Rader, L. V. Yashina, Rapid Surface Oxidation of $Sb_2Te_3$ as Indication for a Universal Trend in the Chemical Reactivity of Tetradymite Topological Insulators, Chem. Mater. 28, 24 (2016) 8916–8923.

[43] A. Audzijonis, L. Žigas, G. Gaigalas, R. Sereika, B. Žygaitienė, Density Functional Calculation of the Photoelectron Emission Spectra of BiSCl Crystal and Molecular Clusters, J. Clust. Sci. 21 (2010) 577–589.

[44] Z. S. Aliev, M. B. Babanly, A. V. Shevelkov, D. M. Babanly, J.-C. Tedenac, Phase diagram of the Sb–Te–I system and thermodynamic properties of SbTeI, Int. J. Mat. Res. 103 (2012) 290–295.

[45] F. D. Murnaghan, The Compressibility of Media under Extreme Pressures, Proc. Natl. Acad. Sci. 30 (1944) 244.

[46] F. Birch, Finite Strain Isotherm and Velocities for Single-Crystal and Polycrystalline NaCl at High Pressures and 300° K, J. Geophys. Res. 83 (1978) 1257.

[47] S. Zhou, J. Long, W. Huang, Theoretical prediction of the fundamental properties of ternary bismuth tellurohalides, Mat. Sci. Semicon. Proc. 27 (2014) 605–610.

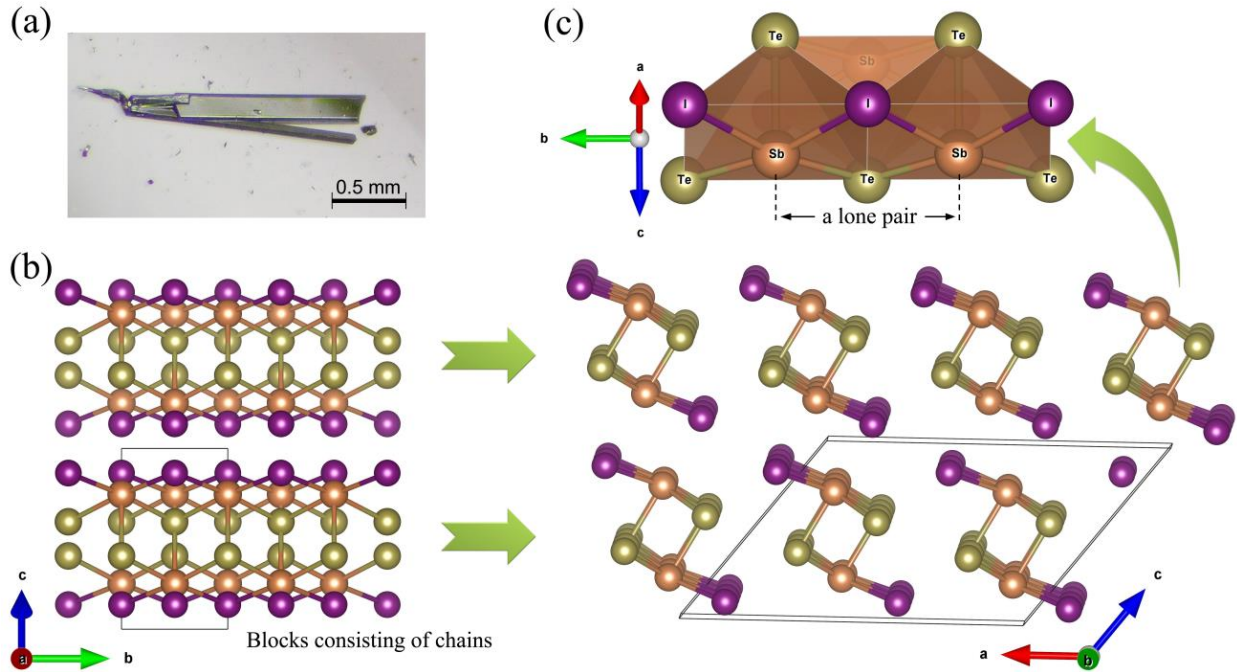

**Figure 1.** Visual and structural view of the SbTeI crystal. (a) Microscope image of the SbTeI single crystal grown by chemical vapor transport. (b) Crystal structure of SbTeI highlighting plate-like blocks and double-chains projected in *c-b* and *a-c* planes, respectively. Here, the double-chains consist of Sb-Te and Sb-I chemical bonds while Te and I atoms are well separated and do not form any connections. The monoclinic unit-cell distinguished by thick solid lines. (c) Coordination polyhedrons of the antimony with a lone pair marked as dashed line.

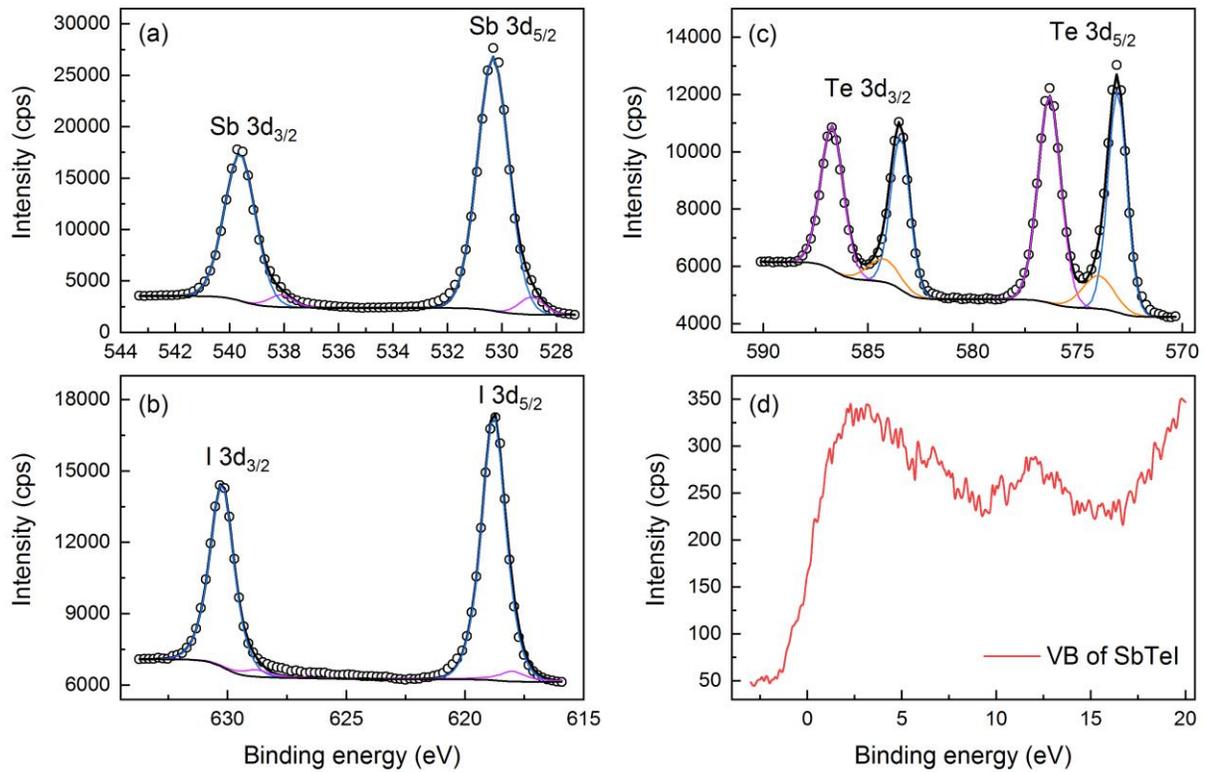

**Figure 2.** XPS spectra of the Sb *3d* (a), I *3d* (b), Te *3d* (c) spin-orbit doublets and valence band (d) collected from the surface of SbTeI crystal.

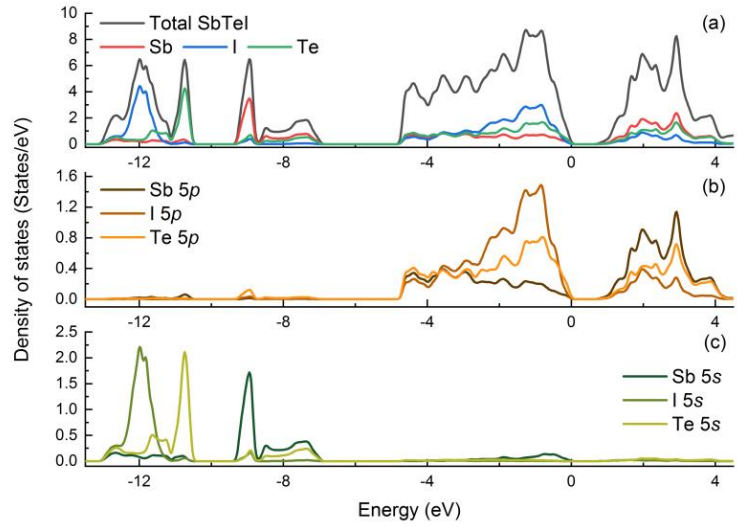

**Figure 3.** The T-DOS (a) and P-DOS (b, c) of SbTeI crystal (monoclinic *C* 2/*m* phase). The valence band has three well separated bands in the energy range from -13.5 to 0 eV. The conduction band situated from 0.7 to 4.2 eV.

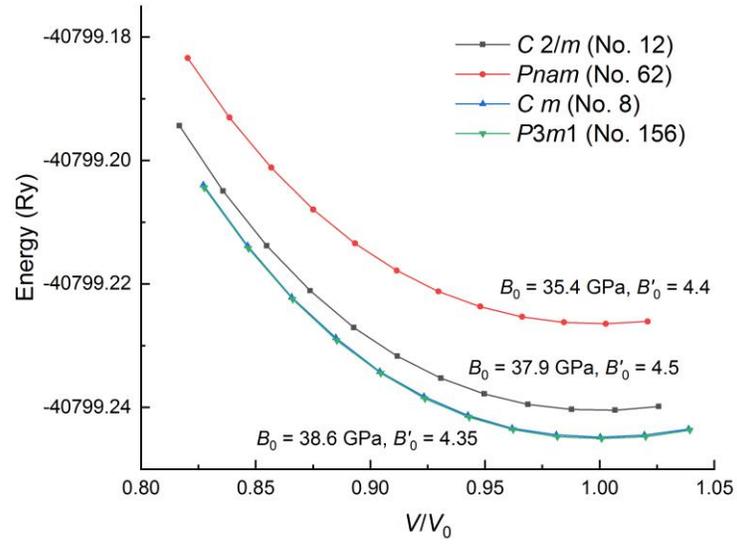

**Figure 4.** Schematic diagram of the energy-volume curves for different phases of SbTeI. The black curve corresponds to the monoclinic *C* 2/*m* phase observed experimentally. The other curves correspond to the theoretically simulated phases described in the text.

**Table 1.** Data collection and structure refinement for SbTeI.

| | |
|---|---|
| Radiation source | Microfocus Incoatec Ims 3.0 (Ag Kα, λ = 0.56086 Å) |
| Theta range for data collection | 2.24 to 22.12° |
| Index ranges | -18<=h<=17, -5<=k<=5, -8<=l<=12 |
| Reflections collected | 2132 |
| Independent reflections | 604 [$R$(int) = 0.0436] |
| Coverage of independent reflections | 100.0% |
| Absorption correction | Multi-Scan |
| Structure solution technique | direct methods |
| Structure solution program | SHELXT 2014/5 (Sheldrick, 2014) |
| Refinement method | Full-matrix least-squares on $F^2$ |
| Refinement program | SHELXL-2018/3 (Sheldrick, 2018) |
| Function minimized | $\Sigma\ w(F_o^2 - F_c^2)^2$ |
| Data / restraints / parameters | 604 / 0 / 19 |
| Goodness-of-fit on $F^2$ | 1.20 |
| Final R indices | 542 data; I>2σ(I) — $R_1$ = 0.0327, $wR_2$ = 0.0750 |
| | all data — $R_1$ = 0.0398, $wR_2$ = 0.0872 |
| Weighting scheme | w=1/[σ²($F_o^2$)+(0.0130P)²+12.5256P] where P=($F_o^2$+2$F_c^2$)/3 |
| Largest diff. peak and hole | 1.792 and -1.711 eÅ$^{-3}$ |
| R.M.S. deviation from mean | 0.387 eÅ$^{-3}$ |
| F(000) | 624 |

**Table 2.** Atomic coordinates and equivalent isotropic atomic displacement parameters (Å$^2$).

| Atom | Space group $C\ 2/m$, $T$ = 296.2 K. | | | | |
|---|---|---|---|---|---|
| | $a$ = 13.698(6) Å, $b$ = 4.2289(18) Å, $c$ = 9.191(4) Å, $\alpha = \gamma$ = 90°, $\beta$ =128.626°, $V$ = 415.9 Å$^3$, $Z$ = 4 | | | | |
| | *x/a* | *y/b* | *z/c* | *U* (eq)* | *Occ.* |
| **Sb** | 0.36588(8) | 1/2 | 0.74674(12) | 0.0271(2) | 1.0 |
| **Te** | 0.32611(8) | 1/2 | 0.40553(12) | 0.0233(2) | 1.0 |
| **I** | 0.58161(8) | 0 | 0.86379(12) | 0.0274(2) | 1.0 |

*$U$(eq) is defined as one third of the trace of the orthogonalized $U_{ij}$ tensor.

**Table 3.** Anisotropic atomic displacement parameters (Å$^2$). The anisotropic atomic displacement factor exponent takes the form: -2π²[h² a*² $U_{11}$ + ... + 2 h k a* b* $U_{12}$].

| | $U_{11}$ | $U_{22}$ | $U_{33}$ | $U_{23}$ | $U_{13}$ | $U_{12}$ |
|---|---|---|---|---|---|---|
| **Sb** | 0.0298(5) | 0.0229(4) | 0.0202(4) | 0 | 0.0116(4) | 0 |
| **Te** | 0.0249(4) | 0.0194(4) | 0.0256(4) | 0 | 0.0158(4) | 0 |
| **I** | 0.0289(5) | 0.0229(4) | 0.0297(5) | 0 | 0.0179(4) | 0 |

**Table 4.** Bond lengths (Å) and angles (°).

| Bond lengths | | | |
|---|---|---|---|
| **Sb-Te** | 2.8281(18) | **Sb-I** (× 2) | 3.2236(15) |
| **Sb-Te** (× 2) | 2.9582(13) | | |
| **Bond angles** | | | |
| **Te-Sb-Te** (× 2) | 88.65(3) | **Sb-Te-Sb** (× 2) | 91.35(3) |
| **Te-Sb-Te** | 91.25(5) | **Sb-Te-Sb** | 91.25(5) |
| **Te-Sb-I** (× 2) | 170.32(4) | **Sb-I-Sb** | 81.98(5) |
| **Te-Sb-I** (× 2) | 82.64(3) | **I-Sb-I** | 81.98(5) |
| **Te-Sb-I** (× 2) | 92.76(4) | | |